\documentclass[english,aps,preprint]{revtex4}
\usepackage[T1]{fontenc}
\usepackage[latin9]{inputenc}
\usepackage{array}
\usepackage{multirow}
\usepackage{amsmath}
\usepackage{graphicx}

\makeatletter

\providecommand{\tabularnewline}{\\}
\newcommand{\lyxdot}{.}

\@ifundefined{textcolor}{}
{%
 \definecolor{BLACK}{gray}{0}
 \definecolor{WHITE}{gray}{1}
 \definecolor{RED}{rgb}{1,0,0}
 \definecolor{GREEN}{rgb}{0,1,0}
 \definecolor{BLUE}{rgb}{0,0,1}
 \definecolor{CYAN}{cmyk}{1,0,0,0}
 \definecolor{MAGENTA}{cmyk}{0,1,0,0}
 \definecolor{YELLOW}{cmyk}{0,0,1,0}
}

\makeatother

\usepackage{babel}
\begin{document}
\title{Excited electron production at the SPPC-based ep colliders via contact
interactions }
\author{A. Caliskan}
\email{acaliskan@gumushane.edu.tr}

\affiliation{Gümü\c{s}hane University, Faculty of Engineering and Natural Sciences,
Department of Physics Engineering, 29100, Gümü\c{s}hane, Türkiye}
\begin{abstract}
If a linear electron accelerator is installed into the SPPC (Super
Proton-Proton Collider) complex, ep collision options will be available
in addition to pp collisions. We consider the production of excited
electrons with spin-1/2 at the future SPPC-based electron-proton colliders
with center-of-mass energies of $8.44$, $11.66$, $26.68$ and $36.88$
TeV. In the $ep\rightarrow e^{\star}X\rightarrow e\gamma X$ signal
process, excited electrons are produced by contact interactions and
decay into the photon channel by gauge interactions. Taking into account
the corresponding background process, the pseudorapidity and transverse
momentum distributions of the final state particles are plotted. We
reported the discovery, observation and exclusion mass limits of excited
electrons by applying appropriate kinematical cuts best suited for
amplify the signal of the excited electron signature. We also investigated
the highest achievable values of the compositeness scale for the discovery
of excited electrons at these colliders.
\end{abstract}
\maketitle

\section{introduction}

The Standard Model (SM) in particle physics is a fundamental theory
that successfully describes the basic particles and three of the four
fundamental interactions between these particles. In $2012$, the
discovery of the Higgs boson at CERN by the ATLAS \cite{observation-ATLAS}
and CMS \cite{observation-CMS} detectors confirmed the Electroweak
Symmetry Breaking mechanism proposed by the SM. With this discovery,
which is a milestone in particle physics, the mechanism of gaining
mass to particles has been experimentally proven. Although all the
experiments performed so far have confirmed the SM, there are many
phenomena that this theory has not yet been able to explain, such
as dark matter, dark energy, elementary particle inflation, family
replication and CP violation. In order to provide a theoretical solution
to these problems, various theories such as Technicolour \cite{weinberg,suskind},
Grand Unified Models \cite{unity of all,lepton number}, Supersymmetry
\cite{extension of the}, Extra Dimensions and Compositeness \cite{souza}
have been proposed. This study has been conducted within the scope
of compositeness theory, as the compositeness can provide a particularly
good explanation for fundamental particle inflation. In these models,
the existence of more fundamental particles called preons has been
proposed. All fermions and their anti-particles are composed of bound
states of the preons. The first studies on the lepton and quark compositeness
began in the $1970s$ \cite{unified model,subquark model,observable effects,a fundamental theory}.
Up to date, numerous preonic models such as Haplon (Fritzsch-Mandelbaum)
\cite{weak interactions,a quantum structure} and Rishon (Harari-Shupe)
\cite{a schematic model,shupe} have been proposed, suggesting new
particles like excited fermions, leptogluons, and leptoquarks within
the scope of these models. According to preonic models, possible new
interactions between fermions occur on the binding energy scale of
the preons. This energy scale, where preons come together to form
SM fermions, is called the compositeness scale and is denoted by $\varLambda$.
If the leptons and quarks in the SM have a composite structure, their
excited states should be observed experimentally as a requirement
of compositeness. Therefore, excited leptons and quarks are among
the proposed new particles. The masses of these proposed particles
are expected to be heavier than their SM counterparts.

In the literature, many studies on excited leptons \cite{excited muon production}\cite{search for excited muons}\cite{single production}\cite{turkish journal of physics}\cite{excited muon searches}\cite{excited neutrino search}
and quarks \cite{excited quarks}\cite{search for excited}\cite{resonance production}
have been carried out for various colliders. In this study, single
production of excited electron by contact interaction method is investigated.
It is a continuation of our previous work \cite{single production of composite}
in which the production by gauge mechanism was investigated. No signal
for the existence of excited leptons has been found in experimental
studies. However, each new study updates the experimental mass limits
of excited leptons. The most recent mass limits for single production
of excited electrons are $3.9$ TeV for gauge decay \cite{mass limit gauge,PDG}
and $5.6$ TeV for contact decay \cite{mass limit contact}. Since
the decay of excited electrons by gauge interactions is considered
in this study, the mass limit of $3.9$ TeV is taken into account.
This mass limit is provided by the CMS detector for the process $pp\rightarrow ee^{\star}X\rightarrow ee\gamma X$,
assuming $f=f'=1$ and $\varLambda=m_{e^{\star}}$.

In this study, the production of excited electrons by contact interactions
and their decay by the gauge mechanism were investigated in the SPPC-based
electron-proton colliders, which is proposed to be established in
China. The rest of the paper is organised as follows: the second section
describes the SPPC-CEPC project and the proposed electron-proton collider
options, the third section discusses the Lagrangian, decay width and
cross sections of excited leptons, the fourth section performs the
signal-background analysis and the last section interprets the results
obtained.

\section{THE SPPC PROJECT AND ELECTRON - PROTON COLLIDERS}

Particle physics has reached the Higgs era with the definitive proof
of the existence of the Higgs particle in experiments conducted at
CERN. In order to further our knowledge on this subject, the properties
of the Higgs particle need to be analysed in more detail. For this
purpose, efforts to establish Higgs factories to produce the Higgs
particle at higher energies have been initiated all over the world.
Thanks to the Higgs factories, more information about the Higgs field
will be obtained by carrying out studies on topics such as the precise
measurement of the Higgs mass and the observation of rare decay products.
Studies on new particles and interactions beyond the SM will also
be carried out at these factories.

Just a few months after the discovery of the Higgs particle, the Chinese
Particle Physics Community proposed the two-stage CEPC-SPPC project.
In the first phase of the project, an electron-positron collider named
Circular Electron Positron Collider (CEPC) will be installed. The
CEPC collider to be built in China will have a tunnel length of approximately
$100$ km, where electron and positron beams travelling in opposite
directions will be collided in detectors to be installed at two points
\cite{cepc accelerator tdr status,cepc AC Power}. The center-of-mass
energies of the collider are targeted to be $91$, $160$ and $240$
GeV with corresponding luminosities of $32$, $10$ and $3x10^{34}cm^{-2}s^{-1}$,
respectively. Since the main purpose of the collider is to investigate
the properties of the Higgs particle, it will work as a Higgs factory
for the first $7$ years and it is expected to produce at least $1$
million Higgs particles during this process. In the following times,
it is planned to produce $1$ trillion Z bosons in $2$ years as a
super Z factory and $100$ million W bosons in $1$ year as a W factory.
In the second phase of the project, a proton-proton collider, the
Super Proton Proton Collider (SPPC), will be built as an energy frontier
and a discovery machine beyond the LHC \cite{CEPC web page}. 

At the SPPC collider, which will share the same tunnel with the CEPC
collider, it will be tried to reach a center-of-mass energy of $70-75$
TeV with dipole magnets of $12$ T in the first stage. The SPPC, which
is expected to reach a luminosity of $1x10^{35}cm^{-2}s^{-1}$, will
be a more powerful $pp$ collider than the LHC at CERN with these
values. At the next stage of the project, it is planned to increase
the center-of-mass energy of the SPPC collider to energies of $125-150$
TeV by using a $20$ T dipole magnets. With these energy values, the
SPPC collider will be more powerful than the $100$ TeV $pp$ collider
in the FCC project \cite{FCC-CDR-1,FCC-CDR-2,FCC-CDR-3,FCC-CDR-4}.
The SPPC collider is planned to be installed after $2040$ \cite{SPPC}.
The Preliminary Content Design Report (Pre-CDR) of the CEPC-SPPC project
was written in $2015$ \cite{SPPC-PreCDR} and the Content Design
Report (CDR) in $2018$ \cite{SPPC-CDR-1,SPPC-CDR-2}. The Technical
Design Report (TDR) of the project is still in progress and is expected
to be finalised in the coming months.

The installation of the CEPC and SPPC colliders in the same tunnel
will enable the $ep$ collision option in addition to $pp$ and $ee^{+}$
collisions. As a result of a preliminary study on this subject, parameters
$\sqrt{s}=4.1$TeV and $L_{ep}=10^{33}cm^{-2}s^{-1}$ were obtained
\cite{preliminary study}. It can be seen that the center-of-mass
energy is quite small here. Because, the problem of synchrotron radiation
in circular electron accelerators prevents reaching high energies.
If a linear electron accelerator is used instead of a circular one,
an $ep$ collider with a higher center-of-mass energy can be obtained.
In another study in this direction, a linear electron accelerator
tangential to the SPPC proton ring was proposed and the basic parameters
for the $ep$ option were derived \cite{SPPC-based}. Higher center-of-mass
energies were achieved in this study, which used the parameters of
the ILC and PWFALC linear electron accelerator projects as the electron
source. For the proton energy, two options were used: $35.6$ TeV,
an energy value that can be reached in the first stage of SPPC, and
$68$ TeV, that can be reached in the second stage. Thus, four different
$ep$ collision options were derived. These parameters are given in
Table $1$.

\begin{table}
\caption{The main parameters of the SPPC - based electron - proton colliders.}

\begin{tabular}{|c|c|c|c|}
\hline 
$E_{e}$$\left(TeV\right)$  & $E_{p}$$\left(TeV\right)$  & $\sqrt{s}$ $\left(TeV\right)$  & $L_{int}$ $\left(cm^{-2}s^{-1}\right)$ \tabularnewline
\hline 
\hline 
$0.5$  & $35.6$  & $8.44$  & $2.51$$\times$ $10^{31}$\tabularnewline
\hline 
$0.5$  & $68$  & $11.66$  & $6.45$$\times$ $10^{31}$ \tabularnewline
\hline 
$5$  & $35.6$  & $26.68$  & $7.37$$\times$ $10^{30}$ \tabularnewline
\hline 
$5$  & $68$  & $36.88$  & $1.89$$\times$ $10^{31}$ \tabularnewline
\hline 
\end{tabular}
\end{table}

\section{GAUGE AND CONTACT INTERACTIONS FOR EXCITED ELECTRONS}

Both the production and decay processes of excited leptons in colliders
can take place by two different interaction mechanisms. If the interaction
between particles is realised by the exchange of specific particles,
this interaction is defined as a gauge interaction. The gauge interaction
Lagrangian between spin-1/2 excited leptons, ordinary leptons and
gauge bosons is given in Equation $1$ \cite{excited lepton-LEP-HERA,excited quark and lepton}.
\begin{center}
\begin{equation}
L_{G}=\frac{1}{2\Lambda}\bar{L_{R}^{\star}}\sigma^{\mu\nu}[fg\frac{\overrightarrow{\tau}}{2}.\overrightarrow{W}_{\mu\nu}+f'g'\frac{Y}{2}B_{\mu\nu}]L_{L}+h.c.,
\end{equation}
\par\end{center}

where $L_{L}$and $L_{R}^{\star}$ denote left-handed ordinary lepton
and right-handed excited lepton, respectively, $\overrightarrow{W}_{\mu\nu}$
and $B_{\mu\nu}$ are the field strength tensors, $\Lambda$ is the
compositeness scale, $f$ and $f'$ are the scaling factors, $g$
and $g'$ are the gauge couplings, Y is hypercharge, $\sigma^{\mu\nu}=i(\gamma^{\mu}\gamma^{\nu}-\gamma^{\nu}\gamma^{\mu})/2$
where $\gamma^{\mu}$ are the Dirac matrices, and $\overrightarrow{\tau}$
represents the Pauli matrices.

The other interaction mechanism of excited leptons is four-fermion
contact interactions which are effective at short distances. The effective
Lagrangian describing this interaction is given in Equation $2$ \cite{excited lepton-LEP-HERA,excited quark and lepton}.
\begin{center}
\begin{equation}
L_{C}=\frac{g_{\star}^{2}}{\Lambda^{2}}\frac{1}{2}j^{\mu}j_{\mu},
\end{equation}
\par\end{center}

\begin{center}
\begin{equation}
j_{\mu}=\eta_{L}\overline{f}_{L}\gamma_{\mu}f_{L}+\eta'_{L}\overline{f^{\star}}_{L}\gamma_{\mu}f_{L}^{\star}+\eta''{}_{L}\overline{f^{\star}}_{L}\gamma_{\mu}f_{L}+h.c.+(L\rightarrow R)
\end{equation}
\par\end{center}

In Equations $2$ and $3$, $g_{\star}$ is the interaction constant
and its value is $g_{\star}^{2}=4\pi$. $\varLambda$ is the compositeness
scale and $j_{\mu}$ represents the left-handed currents. $\eta$
factors are the coefficients of these left-handed currents and their
value is taken as $1$. $f$ and $f^{\star}$ are the SM and the excited
fermion fields, respectively.

When we analyse both Lagrangians, it is seen that the compositeness
scale, $\varLambda$ is inversely proportional. Therefore, it is clearly
seen that as the $\varLambda$ increases for both interactions, the
cross section and decay width values will decrease. However, while
the Lagrangians is inversely proportional to $\varLambda$ in the
gauge interaction, it is inversely proportional to $\varLambda^{2}$
in the contact interaction. This means that gauge interactions dominate
at high $\varLambda$ values and contact interactions dominate at
low $\varLambda$ values. In this study, since the excited electrons
will be produced with the process $e,p\rightarrow e^{\star},j$ by
the contact interaction method, we set $\varLambda$ equal to the
mass of the excited electron, $\varLambda=m_{e^{\star}}$. Thus, we
obtained both a high cross section and a situation in which the contact
interaction is dominant.

Excited electrons can decay by both mechanisms. There are three decay
modes for gauge interactions. These are $e^{\star}\rightarrow e\gamma$,
$e^{\star}\rightarrow eZ$ and $e^{\star}\rightarrow\nu W^{-}$ decay
channels. If we neglect the SM quark masses and for the $m^{\star}>m_{W,Z}$
condition, the analytical formulae giving the decay width of these
channels are given in Equations $4$ below.

\begin{equation}
\varGamma(l^{\star}\rightarrow lV)=\frac{\alpha m^{\star3}}{4\Lambda^{2}}f_{V}^{2}(1-\frac{m_{V}^{2}}{m^{\star2}})^{2}(1+\frac{m_{V}^{2}}{2m^{\star2}}),
\end{equation}

where $V$ represents the $\gamma$, $Z$ and $W^{\pm}$ bosons. $m^{\star}$
is the mass of the excited electron and $m_{v}$ is the mass of the
gauge boson. $f_{V}$ is the interaction constant and its expression
for each decay channel is given in Equation $5$ below.

\[
f_{\gamma}=fT_{3}+f'\frac{Y}{2}
\]

\begin{equation}
f_{Z}=fT_{3}cos^{2}\theta_{W}-f'\frac{Y}{2}sin^{2}\theta_{W},
\end{equation}

\[
f_{W}=\frac{f}{\sqrt{2}}
\]

where $T_{3}$ is the third component of the weak isospin and $Y$
is the hypercharge. For excited electrons, $T_{3}=-\frac{1}{2}$ and
$Y=-1$. $\theta_{W}$is the weak mixing angle.

Three decay channels are also available for contact interactions.
These are the $e^{\star}\rightarrow eq\overline{q}$, $e^{\star}\rightarrow e^{-}e^{-}e^{+}$
and $e^{\star}\rightarrow e\nu\overline{\nu}$ processes, where q
represents quarks. The analytical formula for the decay width of these
processes is given in Equation $6$.

\begin{equation}
\varGamma(l^{\star}\rightarrow lF\overline{F})=\frac{m_{l}^{\star}}{96\pi}(\frac{m_{l}^{\star}}{\Lambda})^{4}N_{C}^{'}S',
\end{equation}

In this Equation, $F$ and $l$ represent SM fermions and leptons,
respectively. $N_{C}^{'}$ is the color factor and has a value of
$1$ for leptons and $3$ for quarks. $S'$ is an additional combinatorical
factor with a value of $1$ if $f\neq l$, $4/3$ for quarks and $2$
for leptons if $f=l$.

For numerical calculations, we implemented both Lagrangians given
in Equations $1$ and $2$ to the CalcHEP simulation package \cite{calchep}
with the help of LanHEP code \cite{lanhep}, and created the model
file for excited electrons. Since we will use the case where $\Lambda$
is equal to $m_{e^{\star}}$ in this study, we calculated the partial
decay widths of both interactions for $\varLambda=m_{e^{\star}}$.
The results are shown in Figure $1$. 

\begin{figure}
\centering{}\includegraphics{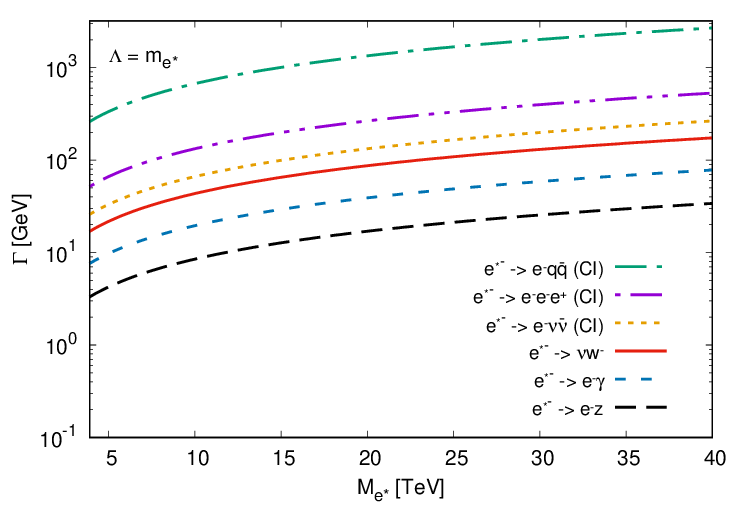}\caption{The partial decay widths for both gauge and contact interactions of
the excited electron for the energy scale $\varLambda=m_{e^{\star}}$}
\end{figure}

When this graph is analysed, it is clearly seen that the decay channels
of the contact interaction are dominant. The $eq\overline{q}$ channel
has the highest decay width values. If we take each of these decay
channels and decay the heavy mass W and Z bosons in the final state,
we obtain the following $6$ possible processes for the ep collider.

\begin{equation}
e^{-}p^{+}\rightarrow e^{\star}p\rightarrow e^{-}q\overline{q}j
\end{equation}

\begin{equation}
e^{-}p^{+}\rightarrow e^{\star}p\rightarrow e^{-}e^{-}e^{+}j
\end{equation}

\begin{equation}
e^{-}p^{+}\rightarrow e^{\star}p\rightarrow e^{-}\nu\overline{\nu}j
\end{equation}

\begin{equation}
e^{-}p^{+}\rightarrow e^{\star}p\rightarrow W^{-}\nu j\rightarrow e^{-}\overline{\nu}\nu j
\end{equation}

\begin{equation}
e^{-}p^{+}\rightarrow e^{\star}p\rightarrow e^{-}\gamma j
\end{equation}

\begin{equation}
e^{-}p^{+}\rightarrow e^{\star}p\rightarrow e^{-}Zj\rightarrow e^{-}q\overline{q}j(e^{-}e^{-}e^{+}j)
\end{equation}

Since the photon channel has fewer final state particles and is easier
to observe in the detector, we chose this channel for this study.
In addition, the photon channel was also used in the last experimental
study conducted by the CMS group, where the most recent mass limits
of excited leptons were determined. In this experimental study, excited
leptons were produced by contact interactions and decayed by gauge
interactions and the photon channel was chosen. Therefore, in this
paper, we have taken exactly this experimental work as an example.
In addition to the mass limits, the CMS group also obtained the most
recent compositeness scale limit. For a $1$ TeV excited lepton mass,
the compositeness scale is excluded up to $25$ TeV.

After selecting the photon decay channel, the cross section values
of the $ep\rightarrow e\gamma j$ process for the proposed $4$ ep
collider options were calculated for $\varLambda=m_{e^{\star}}$ and
the results are shown in Figure $2$. As seen in these graphs, excited
electrons are produced for both interactions and then decayed into
the photon channel. Thus, we have compared both production mechanisms
in these graphs. As it is clearly seen in all graphs, contact production
dominates over gauge production for the $\varLambda=m_{e^{\star}}$
condition. On the other hand, considering the limunosity values in
Table $1$, these four colliders have the capacity to produce a sufficient
number of events. 

\begin{figure}
\begin{centering}
\includegraphics[scale=0.6]{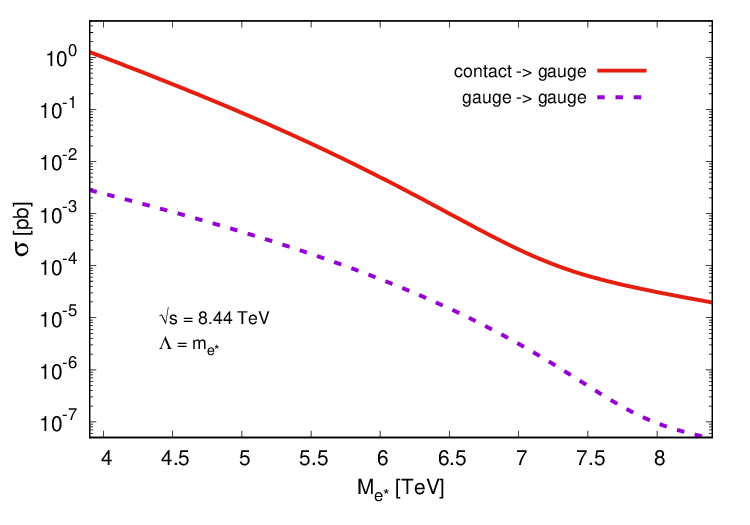}\includegraphics[scale=0.6]{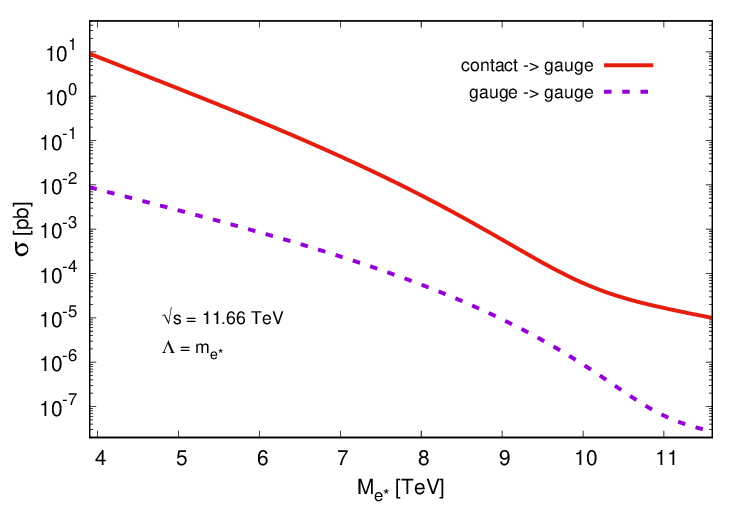}
\par\end{centering}
\begin{centering}
\includegraphics[scale=0.6]{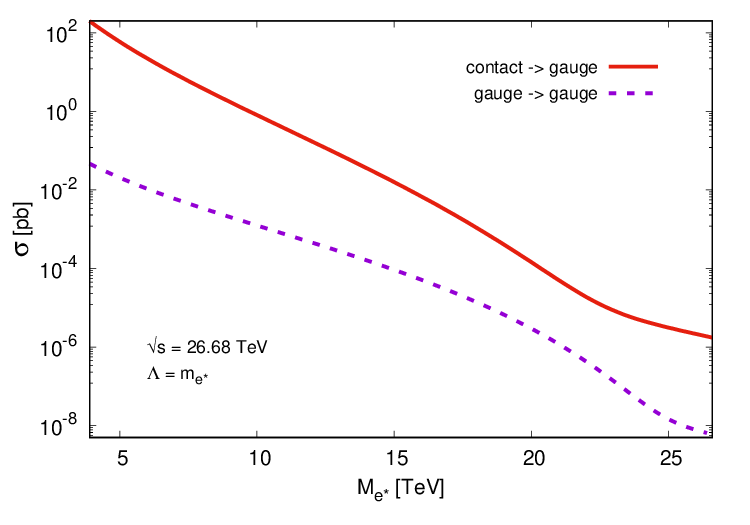}\includegraphics[scale=0.6]{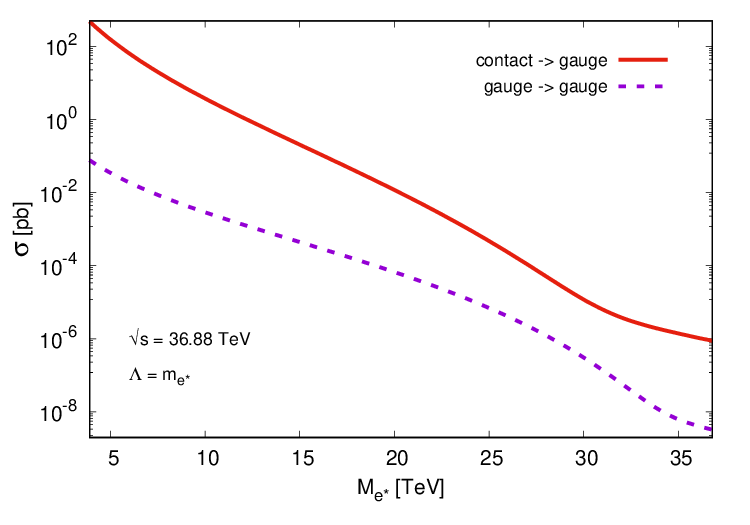}
\par\end{centering}
\caption{The cross-sections of the excited electrons for both production mechanism
with respect to its mass at the SPPC-based electron-proton colliders
with various center-of-mass energies for $\varLambda=m_{e^{\star}}$
and the coupling $f=f'=1$.}

\end{figure}

\section{SIGNAL AND BACKGROUND ANALYSIS}

The SPPC-based $ep$ colliders will enable us to search for excited
electrons through the process $ep\rightarrow e^{\star}j$ (contact
interaction), followed by the subsequent decay of the excited electrons
into an electron and photon (gauge decay). Therefore, our signal process
is $ep\rightarrow e,\gamma,j$, where $j$ represents jets and consists
of quarks and antiquarks. The subprocesses of our signal are of the
form $eq(\overline{q})\rightarrow e\gamma q(\overline{q})$, where
$q$ represents quarks and $\bar{q}$ represents antiquarks. The Feynman
diagram of our signal process is shown in Figure $3$.

\begin{figure}
\begin{centering}
\includegraphics[scale=0.5]{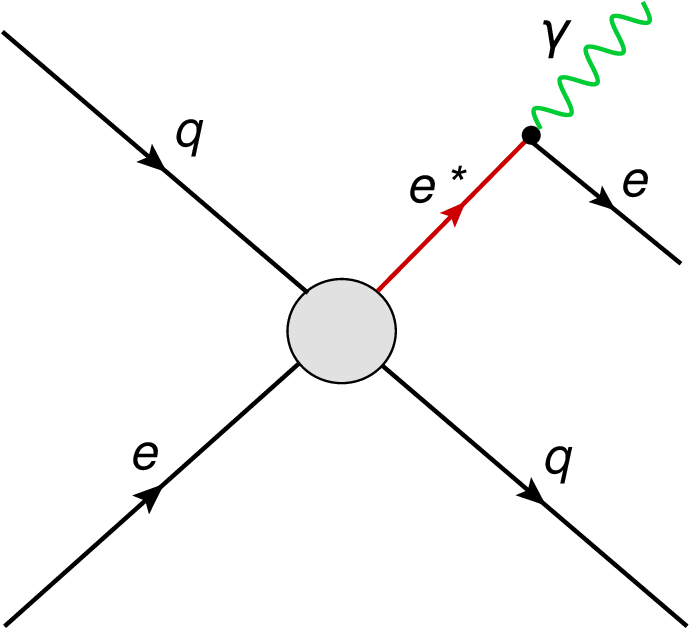}
\par\end{centering}
\caption{Leading-order Feynman diagrams for the signal process $ep\rightarrow e,\gamma,j$. }

\end{figure}

The main background process corresponding to our signal is $ep\rightarrow e,\gamma,j$.
Other complex SM processes are not considered here, because their
contribution would be too small.

\subsection*{Detector Parameters}

Since a detailed detector design for SPPC-based ep colliders has not
yet been carried out, the signal-background analysis is performed
at the parton level. However, in this subsection, the sensitivity
of today's detector technology in the measurement of some basic parameters
will be briefly discussed. For this purpose, the parameters of the
ATLAS detector of the High Luminosity LHC (HL-LHC) project \cite{HL-LHC-TDR},
which is planned to be commissioned in 2029, are considered. The ATLAS
detector is being upgraded to suit the operating conditions of the
HL-LHC. For this purpose, the detector's Inner Tracker system will
be completely replaced with a new silicon-only design. In this way,
it is aimed to achieve higher momentum resolution. In addition, the
pseudorapidity values, which express the tracking range of the detector,
will be extended from $|\eta|<2.5$ to $|\eta|<4$.

In this study, the final state particles of our signal and background
process are electrons, photons and jets. Some important detector parameters
of these particles are reconstruction, identification, isolation efficiency,
energy scale and momentum resolution. The systematic uncertainties
to be achieved at HL-LHC for these parameters are given in Table $2$
\cite{Atlas-parameters}.

\begin{table}

\begin{centering}
\caption{Representative systematic uncertainties in the measurement of some
parameters of electron, photon and jets at the HL-LHC. }
\par\end{centering}
\begin{centering}
\begin{tabular}{|c|c|c|c|}
\hline 
Particles & Parameters & Range & Uncertainty (\%)\tabularnewline
\hline 
\hline 
\multirow{4}{*}{Electron} & Energy Scale & $P_{T}\approx45$GeV & $0.1$ \tabularnewline
\cline{2-4} \cline{3-4} \cline{4-4} 
 & Energy Scale & up to $200$ GeV & $0.3$\tabularnewline
\cline{2-4} \cline{3-4} \cline{4-4} 
 & Reconstruction+Identification Efficiency (ID) & $P_{T}\approx45$GeV & $0.5$\tabularnewline
\cline{2-4} \cline{3-4} \cline{4-4} 
 & Reconstruction+ID+Isolation Efficiency & $P_{T}>200$GeV & $2$\tabularnewline
\hline 
\multirow{4}{*}{Photon} & Energy Scale & $P_{T}\approx60$GeV & $0.3$\tabularnewline
\cline{2-4} \cline{3-4} \cline{4-4} 
 & Energy Scale & up to $200$ GeV & $0.5$\tabularnewline
\cline{2-4} \cline{3-4} \cline{4-4} 
 & Resolution & $P_{T}\approx60$GeV & $10$\tabularnewline
\cline{2-4} \cline{3-4} \cline{4-4} 
 & Reconstruction+ID+Isolation Efficiency & $P_{T}<200$GeV & $2$\tabularnewline
\hline 
\multirow{8}{*}{Jets} & Absolute Jet Energy Scale & - & $1-2$\tabularnewline
\cline{2-4} \cline{3-4} \cline{4-4} 
 & Pileup & - & $0-2$\tabularnewline
\cline{2-4} \cline{3-4} \cline{4-4} 
 & Jet Flavour Composition & - & $0-0.5$\tabularnewline
\cline{2-4} \cline{3-4} \cline{4-4} 
 & Jet Flavour Response & - & $0-0.8$\tabularnewline
\cline{2-4} \cline{3-4} \cline{4-4} 
 & b-jet efficiency & $30<P_{T}<300$ GeV & $1$\tabularnewline
\cline{2-4} \cline{3-4} \cline{4-4} 
 & b-jet efficiency & $P_{T}>300$ GeV & $2-6$\tabularnewline
\cline{2-4} \cline{3-4} \cline{4-4} 
 & c-jet efficiency & all working points & $2$\tabularnewline
\cline{2-4} \cline{3-4} \cline{4-4} 
 & light-jet mistag & working-point dependent & $5-15$\tabularnewline
\hline 
\end{tabular}
\par\end{centering}
\end{table}

According to the data in this table, it is understood that very high
precision measurements can be made. In a future ep detector, with
the further development of technology, much more sensitive measurements
will be possible.

\subsection*{Kinematical Cuts For Discovery Of Excited Electrons}

In the signal-background analysis, we first applied pre-selection
cuts $P_{T}^{e,\gamma,j}>25$ GeV to the transverse momentum of the
final state particles, electron, photon and jet in order to separate
the excited electron signals from the background. We then obtained
some kinematic distributions for both signal and background and superimposed
them on the same graph, so that we could compare signal and background.
First, transverse momentum ($P_{T}$) and pseudorapidity ($\eta$)
distributions for the final state particles, electrons, photons and
jets, of the ep collider with a center-of-mass energy of $8.44$ TeV
are plotted. In these distributions, mass values of $m_{e^{\star}}=3900,$$5000$,
$6000$, $7000$ GeV were used for the signal. Since the $P_{T}$
and $\eta$ distributions of the electron and photon are similar,
only the distributions of the electron are given and these are shown
in Figure $4$.

\begin{figure}
\begin{centering}
\includegraphics[scale=0.6]{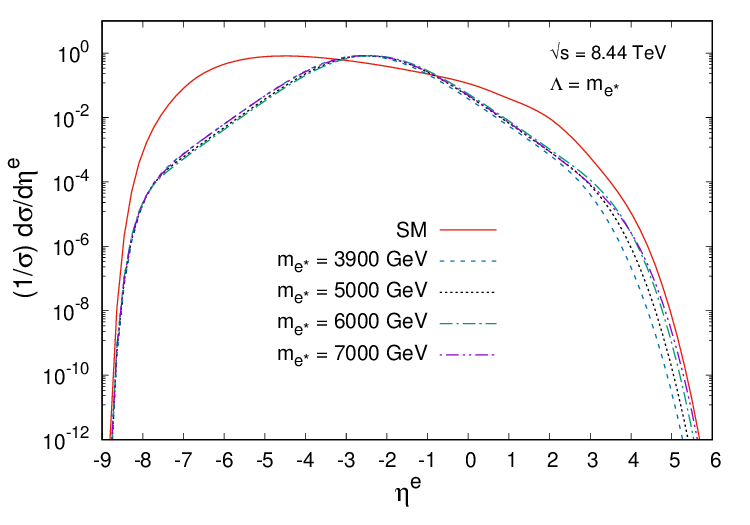}\includegraphics[scale=0.6]{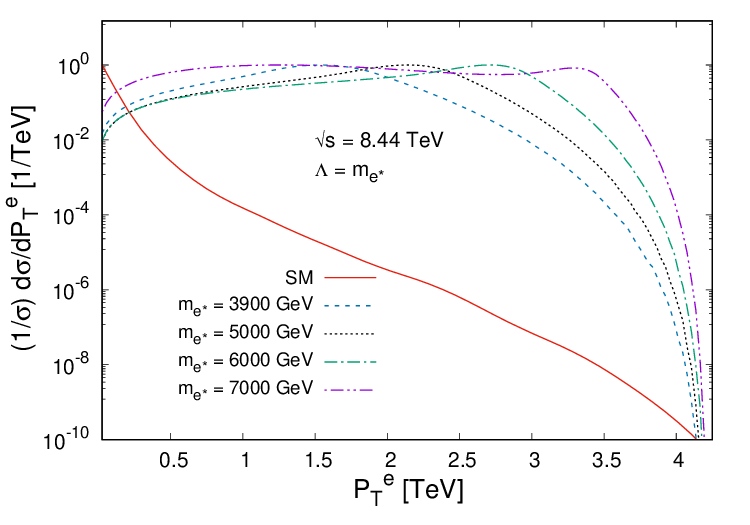}
\par\end{centering}
\caption{The normalized pseudorapidity (left) and transverse momentum (right)
distributions of the final state electrons for the ep collider with
center-of- mass energy $8.44$ TeV.}

\end{figure}

Similar procedures were performed for the other ep colliders with
center-of-mass energies of $11.66$, $26.68$ and $36.88$ TeV and
the resulting distributions are shown in Figure $5$, Figure $6$
and Figure $7$, respectively.

\begin{figure}
\begin{centering}
\includegraphics[scale=0.6]{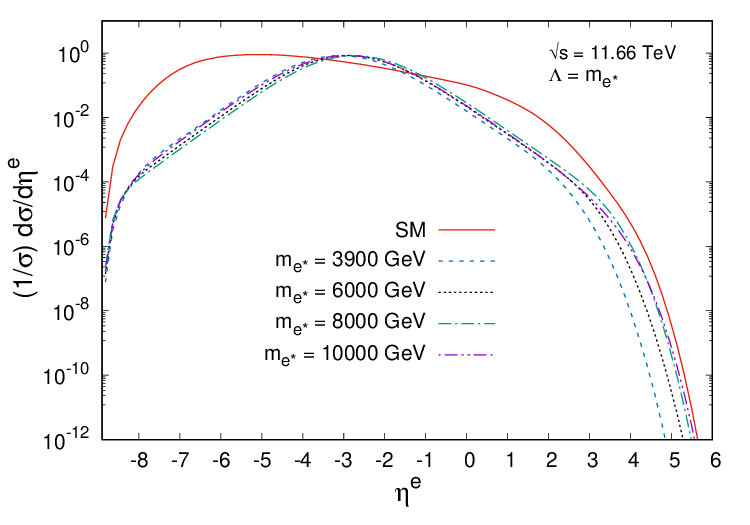}\includegraphics[scale=0.6]{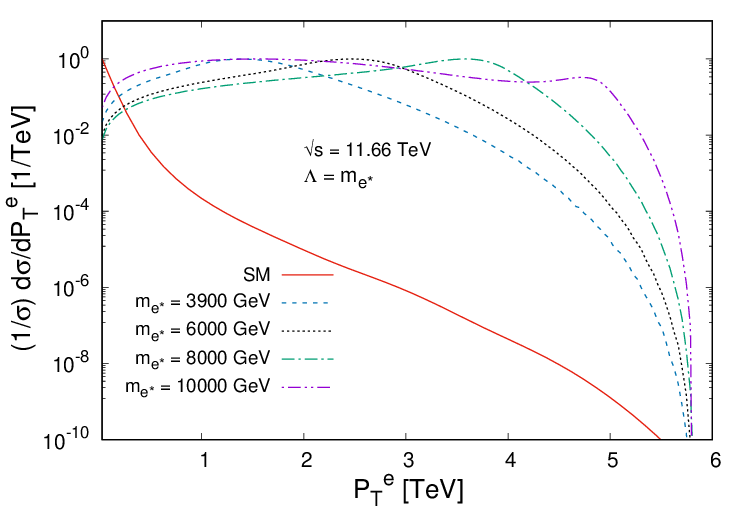}
\par\end{centering}
\caption{The normalized pseudorapidity (left) and transverse momentum (right)
distributions of the final state electrons for the ep collider with
center-of- mass energy $11.66$ TeV.}

\end{figure}

\begin{figure}
\begin{centering}
\includegraphics[scale=0.6]{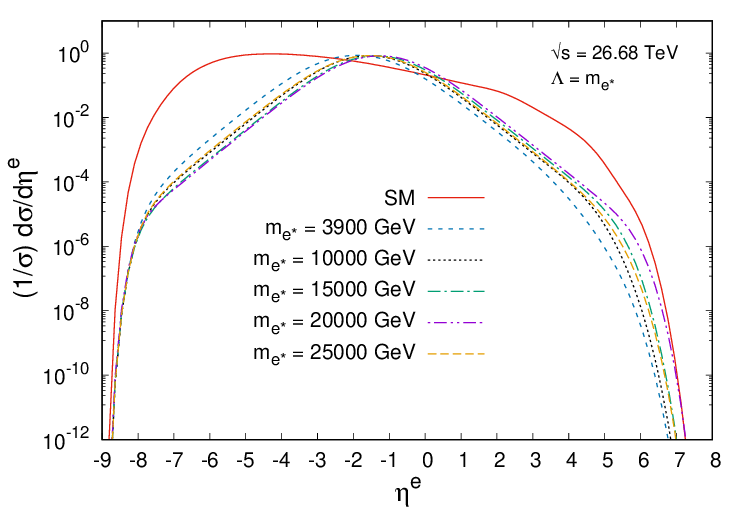}\includegraphics[scale=0.6]{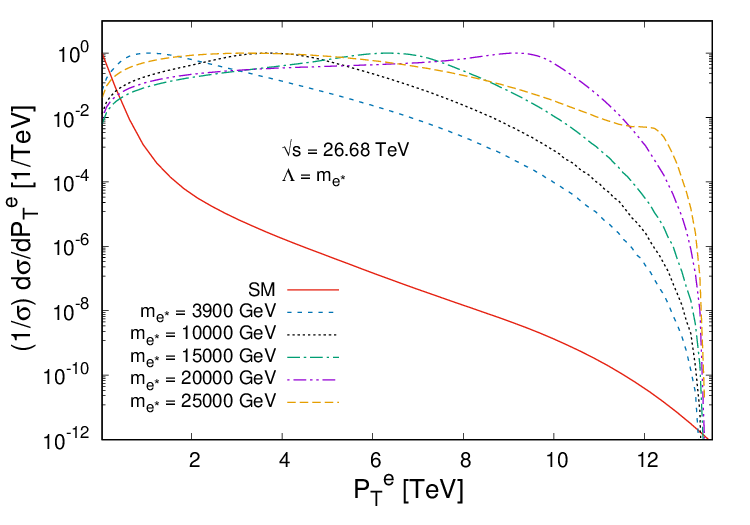}
\par\end{centering}
\caption{The normalized pseudorapidity (left) and transverse momentum (right)
distributions of the final state electrons for the ep collider with
centre-of- mass energy $26.68$ TeV.}

\end{figure}

\begin{figure}
\begin{centering}
\includegraphics[scale=0.6]{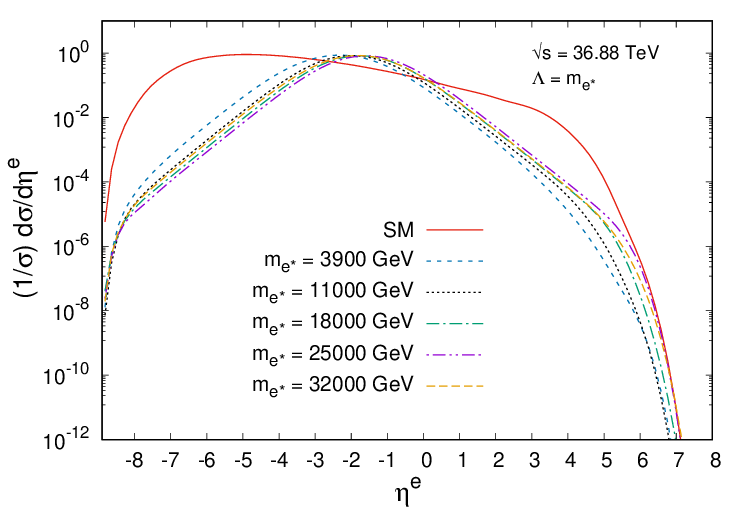}\includegraphics[scale=0.6]{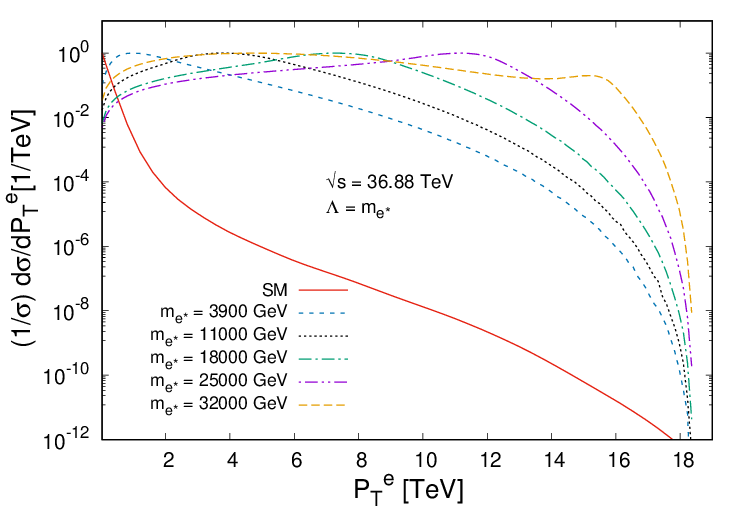}
\par\end{centering}
\caption{The normalized pseudorapidity (left) and transverse momentum (right)
distributions of the final state electrons for the ep collider with
centre-of- mass energy $36.88$ TeV.}

\end{figure}

When we examine the pseudorapidity plots of the final state particles,
it is clearly seen that these distributions peak in the negative region
at all ep colliders. Considering that pseudorapidity is mathematically
defined as $\eta=-\ln\tan(\theta/2)$, where $\theta$ is polar angle,
it is understood that electrons and photons are spatially backward,
so we can say that excited electrons are mostly produced in the backward
direction. This is mainly due to the asymmetric nature of the ep colliders.
Since the energy of the electron is smaller, the pseudorapidity distributions
are boosted towards the side from which the electron beam comes. Therefore,
they peaked in the negative region.

On the other hand, when all $P_{T}$ and $\eta$ plots are analysed,
it is seen that the signal and background distributions are slightly
separated from each other. However, since the cross section of the
background is larger, this separation is not sufficient to identify
the signal from the background. Therefore, in addition to the pre-selection
cuts, we need to apply large cuts, so-called discovery cuts. If we
select regions $-3.5<\eta^{e}<-0.5$ in the $\eta$ plot and $p_{T}^{e}>500$
GeV in the $P_{T}$ plots in Figure $4$, these cuts will hardly change
the cross section of the signal. On the other hand, they will dramatically
reduce the cross section of the background. A similar method was followed
for the other $P_{T}$ and $\eta$ distributions, i.e. discovery cuts
were determined so as not to affect the signal too much and to reduce
the background. Determined discovery cuts for the $P_{T}$ and $\eta$
distributions of the particles of electron, photon and jet in the
final state are reported in Table $3$. This table also shows the
discovery cuts of the jets. But, their distribution plots are not
given in this paper since the jets are not directly related to our
signal.

\begin{table}
\caption{The discovery cuts in the $P_{T}$ and $\eta$ distributions of final
state particles at SPPC-based ep colliders.}

\centering{}%
\begin{tabular}{|c|c|c|c|c|c|c|}
\hline 
$\sqrt{s}$ {[}TeV{]} & $p_{T}^{e}$ & $p_{T}^{\gamma}$ & $p_{T}^{j}$ & $\eta^{e}$ & $\eta^{\gamma}$ & $\eta^{j}$\tabularnewline
\hline 
\hline 
$8.44$ & $p_{T}^{e}$ > $500$ GeV & $p_{T}^{\gamma}$ > $500$ GeV & $p_{T}^{j}$ > $500$ GeV & -$3.5$ < $\eta^{e}$< -$0.5$ & -$3.5$ < $\eta^{\gamma}$< -$0.5$ & -$4$ < $\eta^{j}$< $2.5$\tabularnewline
\hline 
$11.66$ & $p_{T}^{e}$ > $500$ GeV & $p_{T}^{\gamma}$ > $500$ GeV & $p_{T}^{j}$ > $500$ GeV & -$3.5$ < $\eta^{e}$< -$1$ & -$3.5$ < $\eta^{\gamma}$< -$1$ & -$4$ < $\eta^{j}$< $2.5$\tabularnewline
\hline 
$26.68$ & $p_{T}^{e}$ > $500$ GeV & $p_{T}^{\gamma}$ > $500$ GeV & $p_{T}^{j}$ > $500$ GeV & -$2.5$ < $\eta^{e}$< $0.5$ & -$2.5$ < $\eta^{\gamma}$< $0.5$ & -$4$ < $\eta^{j}$< $2.5$\tabularnewline
\hline 
$36.88$ & $p_{T}^{e}$ > $500$ GeV & $p_{T}^{\gamma}$ > $500$ GeV & $p_{T}^{j}$ > $500$ GeV & -$2.5$ < $\eta^{e}$< $0$ & -$2.5$ < $\eta^{\gamma}$< $0$ & -$4$ < $\eta^{j}$< $2.5$\tabularnewline
\hline 
\end{tabular}
\end{table}

One of the most powerful methods to separate the signal from the background
is to apply a cut to the electron-photon invariant mass distributions.
The invariant mass distribution plots obtained after applying pre-selection
cuts are shown in Figure $8$. In these plots, it can be seen that
the line belonging to the background distribution is below the signal
peaks. If we apply an invariant mass cut in the form $m_{e^{\star}}-2\varGamma_{e^{\star}}<m_{e\gamma}<m_{e^{\star}}+2\varGamma_{e^{\star}}$,
where $\varGamma$ shows the decay width of excited electron, we can
make this line lower. The invariant mass cut is much more effective
than the others. So we also applied this effective cut.

\begin{figure}
\begin{centering}
\includegraphics[scale=0.6]{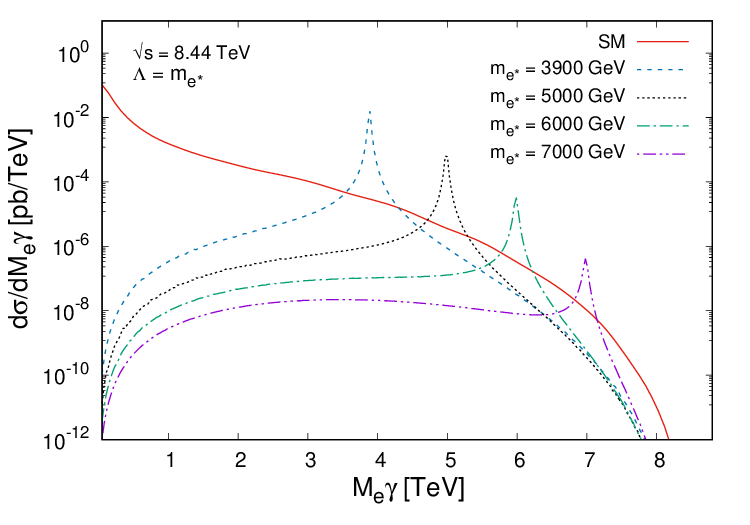}\includegraphics[scale=0.6]{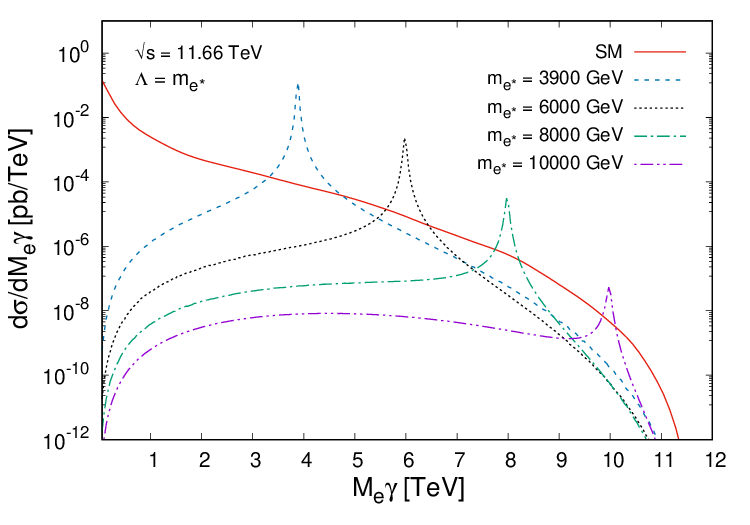}
\par\end{centering}
\begin{centering}
\includegraphics[scale=0.6]{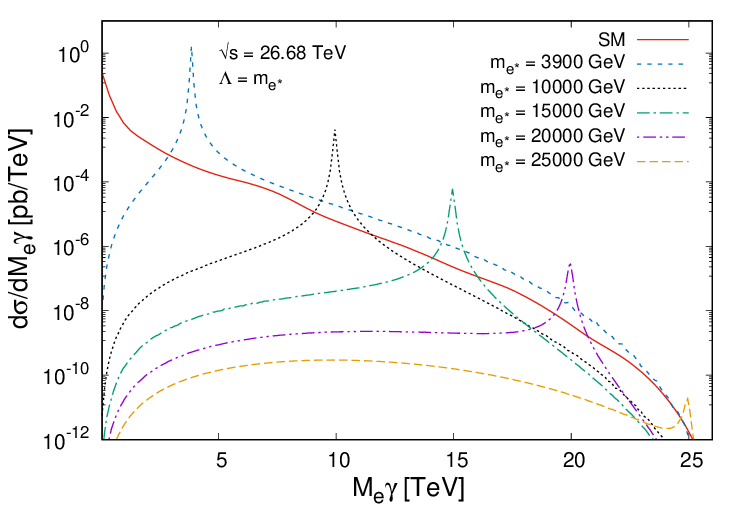}\includegraphics[scale=0.6]{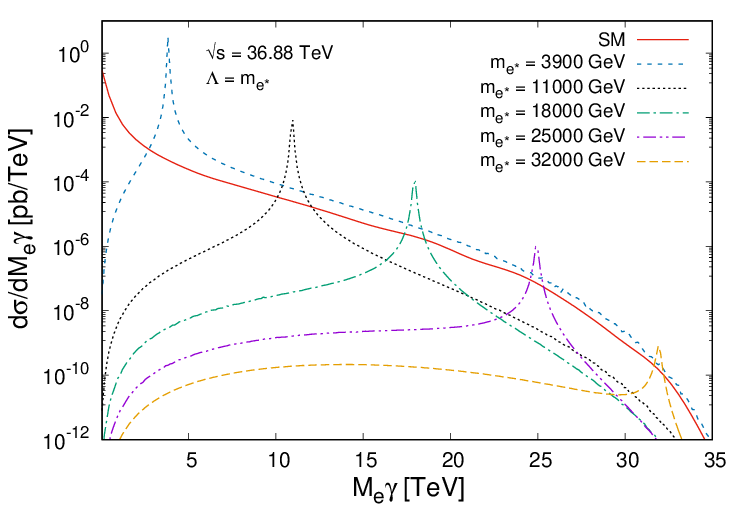}
\par\end{centering}
\caption{The invariant mass distributions of excited electron and corresponding
background for SPPC-based ep colliders.}

\end{figure}

In addition to the above mentioned cuts, we applied some separation
cuts in order to distinguish the final-state particles from each other.
We applied the $\Delta R(e,\gamma)=0.7$ \cite{mass limit gauge}
cut to separate the electron from the photon and the $\Delta R(j,\gamma)=\Delta R(j,e)=0.4$
\cite{mass limit contact} cuts to separate the jets from the electron
and photon. Here $\Delta R$ is the separation cut and is defined
as $\Delta R=\sqrt{\Delta\eta^{2}+\Delta\phi^{2}}$.

\subsection*{Significance Calculus}

In the signal-background analysis, the discovery cuts mentioned in
the previous subsection were used to separate the signal from the
background and the Statistical Significance (SS) values of the expected
signal yield were calculated. The following formula was used to calculate
the SS values \cite{SS}.

\begin{equation}
SS=\sqrt{2\left[\left(S+B\right)\ln\left(1+\left(\frac{S}{B}\right)\right)-S\right]},
\end{equation}

where S and B denote event numbers of the signal and background, respectively.
Using the data obtained as a result of the calculations, the variation
graphs of the SS with respect to the mass of the excited electron
were plotted. SS plots for all colliders are shown in Figure 9.

\begin{figure}
\begin{centering}
\includegraphics[scale=0.6]{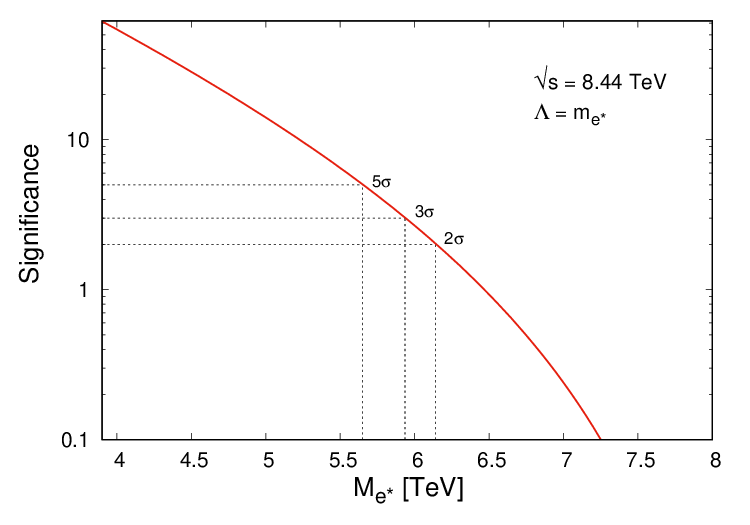}\includegraphics[scale=0.6]{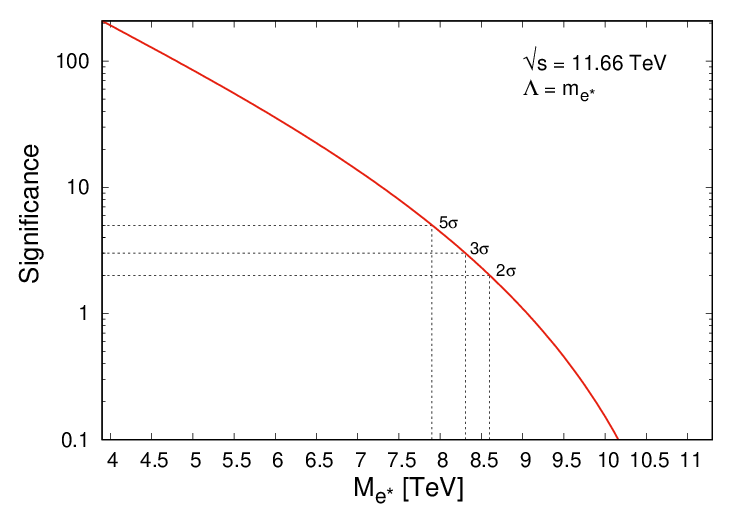}
\par\end{centering}
\begin{centering}
\includegraphics[scale=0.6]{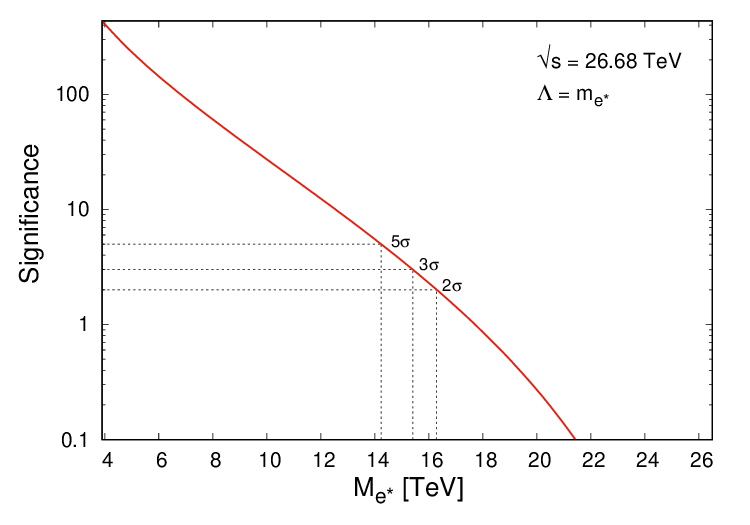}\includegraphics[scale=0.6]{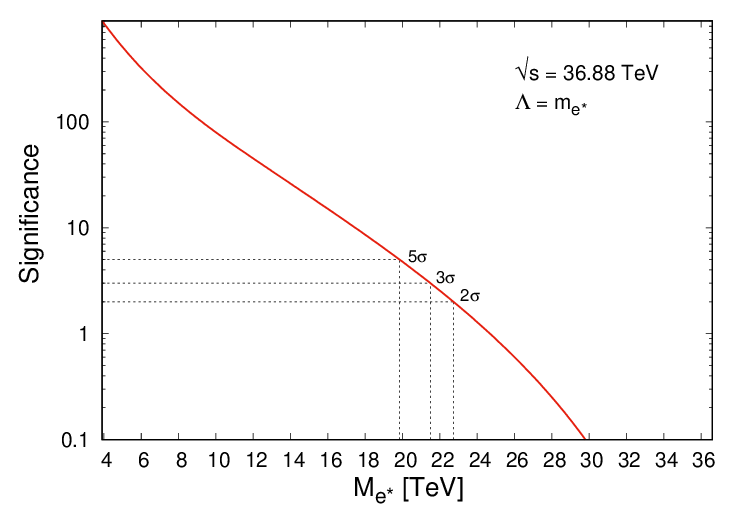}
\par\end{centering}
\caption{Plots of the variation of statistical significance (SS) with respect
to the mass of the excited electron at SPPC-based ep colliders}

\end{figure}

Afterwards, more detailed calculations were performed to obtain the
discovery ($5\sigma$), observation ($3\sigma$) and exclusion ($2\sigma$)
values of the mass of the excited electron. According to the findings,
the ep collider with a center-of-mass energy of $8.44$ TeV will have
the potential to discover excited electrons up to mass $5650$ GeV,
observe them up to mass $5935$ GeV and exclude them up to mass $6140$
GeV. The mass limits obtained for all colliders are reported in Table
$4$. 

\begin{table}
\caption{Attainable mass limits of the excited electrons for SPPC-based ep
colliders}

\begin{centering}
\begin{tabular}{|c|c|c|c|}
\hline 
$\sqrt{s}$ (TeV) & $5\sigma$ (GeV) & $3\sigma$ (GeV) & $2\sigma$ (GeV)\tabularnewline
\hline 
\hline 
$8.44$ & $5650$ & $5935$ & $6140$\tabularnewline
\hline 
$11.66$ & $7900$ & $8305$ & $8600$\tabularnewline
\hline 
$26.68$ & $14230$ & $15410$ & $16290$\tabularnewline
\hline 
$36.88$ & $19840$ & $21485$ & $22725$\tabularnewline
\hline 
\end{tabular}
\par\end{centering}
\end{table}

In addition to these calculations, the highest compositeness scale
values that can be achieved for each collider were calculated. The
calculations show that as the mass of the excited electron increases,
the compositeness scale values decrease inversely. Therefore, in order
to reach the highest compositeness scale values, we should look for
the smallest mass value. Considering that excited electrons are experimentally
excluded up to $3.9$ TeV, it would be appropriate to take the mass
of the excited electron as $4$ TeV for this calculation. Choosing
the mass of the excited electron at $4$ TeV, the values of the compositeness
scale corresponding to $2\sigma$, $3\sigma$ and $5\sigma$ were
calculated for each ep collider and the results are listed in Table
$5$. According to these results, the highest compositeness scale
value can be reached at the collider with a center-of-mass energy
of $36.88$ TeV. At this collider, the compositeness scale for the
discovery of the excited electron is $41915$ GeV.

\begin{table}
\caption{Attainable compositeness scale limits of the excited electrons with
a mass of 4 TeV for SPPC-based ep colliders}

\begin{centering}
\begin{tabular}{|c|c|c|c|}
\hline 
$\sqrt{s}$ (TeV) & $5\sigma$ (GeV) & $3\sigma$ (GeV) & $2\sigma$ (GeV)\tabularnewline
\hline 
\hline 
$8.44$ & $11800$ & $14780$ & $17625$\tabularnewline
\hline 
$11.66$ & $21000$ & $26200$ & $31150$\tabularnewline
\hline 
$26.68$ & $29830$ & $37215$ & $44210$\tabularnewline
\hline 
$36.88$ & $41915$ & $52260$ & $62050$\tabularnewline
\hline 
\end{tabular}
\par\end{centering}
\end{table}

\section{CONCLUSION}

In this study, we investigate the production of excited electrons
by contact interactions and their decay into the photon channel by
gauge interactions at ep colliders. Calculations were performed for
four different SPPC-based electron-proton colliders with center-of-mass
energies of $8.44$, $11.66$, $26.68$ and $36.88$ TeV. In the signal-background
analysis, in addition to pre-selection cuts, discovery cuts were applied
to separate the excited electron signal from the background. In all
calculations for the signal, the compositeness scale was taken equal
to the mass of the excited electron. According to the results, excited
electrons can be discovered up to $5650$ GeV at a collider with a
center-of-mass energy of $8.44$ TeV and an integrated luminosity
of 251 $pb^{-1}$, and up to $7900$ GeV at a collider with a center-of-mass
energy of $11.66$ TeV and an integrated luminosity of $645$ $pb^{-1}$.
In the last two high-energy ep colliders, the collider with a center-of-mass
energy of $26.68$ TeV and an integrated luminosity of $73.7$ $pb^{-1}$
will be able to discover up to $14230$ GeV, and the collider with
a center-of-mass energy of $36.88$ TeV and an integrated luminosity
of $189$ $pb^{-1}$ will be able to discover up to $19840$ GeV. 

In the last part of the study, the highest compositeness scale achievable
at these colliders was investigated. Accordingly, excited electrons
with a mass of $4$ TeV can be discovered up to $11800$ GeV at the
ep collider with a center of mass energy of $8.44$ TeV, $21000$
GeV at the collider with a center of mass energy of $11.66$ TeV,
$29830$ GeV at the collider with a center of mass energy of $26.68$
TeV and $41915$ GeV at the collider with a center of mass energy
of $36.88$ TeV. All these calculations for excited electrons show
that SPPC-based ep colliders will provide the possibility to scan
a wide mass range for excited lepton searches. The observation of
any excited lepton signal at these colliders will provide direct evidence
for the existence of composite models.$ $

\subsection*{Conflict of Interest}

The author declares that he has no conflict of interest.
\begin{acknowledgments}
I would like to thank Dr. M. Sahin from Usak University, Türkiye,
for his support for the model file and for the useful consultations
we had.
\end{acknowledgments}

\end{document}